\documentclass[aps]{revtex4}
\usepackage{graphicx} 
\usepackage{amsmath,amssymb,subfigure}
\usepackage{color}
\def\bc{\begin{center}}
\def\ec{\end{center}}

\def\beq{\begin{equation}}
\def\eeq{\end{equation}}

\begin{document}

\newcommand{\mcKp}{\mathcal{K}\texttt{e}^{i\Theta}}
\newcommand{\mcKm}{\mathcal{K}\texttt{e}^{-i\Theta}}
\newcommand{\p}{\partial}

\title{Superfluidity of dipole excitons in two layers of gapped graphene}

\author{Oleg L. Berman$^{1,2}$, Roman Ya. Kezerashvili$^{1,2}$, and Klaus G. Ziegler$^{3}$}
\affiliation{\mbox{$^{1}$Physics
Department, New York City College
of Technology, The City University of New York,} \\
Brooklyn, NY 11201, USA \\
\mbox{$^{2}$The Graduate School and University Center, The
City University of New York,} \\
New York, NY 10016, USA \\
\mbox{$^3$   Institut f\"ur Physik, Universit\"at Augsburg\\
D-86135 Augsburg, Germany }}

\date{\today}

\begin{abstract}
 A study of the formation of excitons as a problem of two Dirac particles confined
 in two-layer graphene sheets separated by a dielectric when gaps are opened and
 they interact  via a Coulomb potential is presented. We propose to observe Bose-Einstein
condensation and superfluidity of quasi-two-dimensional dipole
excitons in double layer graphene in the presence of band gaps. The
energy spectrum of the collective excitations, the sound spectrum,
and the effective exciton mass are functions of the energy gaps,
density and interlayer separation.  The superfluid density $n_{s}$
and temperature of the Kosterlitz-Thouless phase transition $T_{c}$
are decreasing functions of the energy gaps as well as the
interlayer separation, and therefore, could be controlled by these
parameters.
\end{abstract}

\maketitle






\section{Introduction}

\label{intro}

The many-particle systems of the spatially-indirect dipole excitons
in coupled quantum wells (CQW's)  have been the subject of recent
experimental investigations \cite{Snoke,Butov,Timofeev,Eisenstein}.
These systems are of interest, in particular, in connection with the
possibility of Bose-Einstein condensation (BEC) and
superfluidity of dipole excitons or electron-hole pairs, which would
manifest itself in the CQW's as persistent electrical currents in each
well and also through coherent optical properties and Josephson
phenomena \cite{Lozovik,Littlewood,Vignale,Berman}.

Recent technological advances have allowed the production of
graphene, which is a 2D honeycomb lattice of carbon atoms that form
the basic planar structure in graphite \cite{Novoselov1,Zhang1}.
Graphene has been attracting a great deal of experimental and
theoretical attention because of its unusual properties in its band
structure \cite{Novoselov2,Zhang2,Nomura,Jain}. It is a gapless
semiconductor with massless electrons and holes which have been
described as Dirac-fermions \cite{DasSarma}. Due to the absence of a
gap between the conduction and valence bands in graphene, the
screening effects result in the absence of excitons in graphene.
However, the gap in the electron spectrum in graphene can be opened
by applying the magnetic field, which results in the formation of
magnetoexcitons~\cite{iyengar07}. The BEC and superfluidity of
spatially-indirect magnetoexcitons with spatially separated
electrons and holes in high magnetic field have been studied in
graphene double layer \cite{Berman_L_G} and graphene superlattice
\cite{Berman_K_L,Berman_K_L_2}. The electron-hole pair condensation
in the graphene-based bilayers have been studied
in~[\onlinecite{Sokolik,MacDonald1,MacDonald2,Efetov}]. However, the
effective mass of magnetoexcitons increases when the magnetic field
increases and, therefore, the Kosterlitz-Thouless critical
temperature of the superfluidity decreases with increasing
 magnetic field.

In this paper we propose a new physical realization of an excitonic
BEC and superfluidity in two parallel graphene layers, when one layer
 is filled by electrons, and the other one is filled by holes. We consider two parallel
graphene layers separated by an insulating slab   (e.g. $SiO_2$) and propose the formation of the excitons
due to the gap opening in the electron and hole spectra in the two graphene layers. The
advantage of the consideration of exciton formed by an electron and a hole
from two different graphene layers, separated by an insulating slab, is that the dielectric
slab creates the barrier for the electron-hole recombination which increases the life-time of the
exciton compared to the exciton formed by an electron and a hole in a single
graphene layer.
\par
There are different mechanisms of the band gap
opening in graphene. Substrate-induced band gap opening in epitaxial
graphene is caused by the breaking of sublattice symmetry owing to the
graphene substrate interaction~\cite{Zhou}. When graphene is epitaxially
grown on SiC substrate, a gap of $\sim 0.26 \ \mathrm{eV}$ is
produced~\cite{Zhou}. The electronic structure of graphene can be
tuned by an organic molecule. The band gap can be opened in graphene
due to the charge transfer between an organic molecule and
graphene~\cite{Lu}. It was demonstrated  by angle-resolved
photoemission spectroscopy that a tunable gap in quasi-free-standing
monolayer graphene on Au can be induced by
hydrogenation~\cite{Haberer}. The size of the gap can be controlled
by hydrogen loading and reaches $\sim 1.0 \ \mathrm{eV}$ for a
hydrogen coverage of $8 \mathrm{\%}$~\cite{Haberer}. The band gap
tuning in hydrogenated graphene was also analyzed within the density
functional theory~\cite{Gao}.

The equilibrium system of local pairs of spatially
separated electrons and holes
can be created by varying the chemical potential, using a
bias voltage between two graphene layers or between two gates
located near the corresponding graphene sheets. For
simplicity, we also call these equilibrium local electron-hole pairs
as indirect excitons. Excitons with spatially separated electrons
and holes can be created also by laser pumping (far infrared in
graphene) and by applying perpendicular electric field as
for the CQW's \cite{Snoke,Butov,Eisenstein}.
We assume that the system is in a quasi-equilibrium state.
Below we study the low-density regime for excitons, i.e. exciton radius $a < n^{-1/2}$, where $n$ is the 2D exciton density.
In this system the effective exciton mass can be controlled by the
gap. The effective exciton mass can be small relative to the mass of
free electron, and the Kosterlitz-Thouless transition temperature
$T_{c}$ controlled by the gap is expected to be the same order or  relatively high
compared to the coupled quantum wells case.

Our paper is organized in the following way. In Sec.~\ref{ham} we
present the Hamiltonian of the spatially separated electron and hole in two different parallel graphene sheets separated by a dielectric
in the presence of the band gap.  In Sec.~\ref{sol}
 we obtain the single-particle energy spectrum of dipole excitons in two-layers graphene and find the effective exciton mass. In Sec.~\ref{col} we obtain
  the spectrum of collective excitation in the weakly-interacting gas of dipole excitons. The density of the superfluid component
  and the temperature of the phase transitions for the system of dipole excitons in two-layer graphene in the presence of a band gap are obtained in Sec.~\ref{sup}.
  Finally, the discussion of the results and conclusions follow in Sec.~\ref{disc}.


\section{Exciton Hamiltonian}
\label{ham}

Let us consider a system of electrons and holes located in two different parallel graphene sheets. In this system electrons and holes move in two separate sheets with honeycomb lattice structure. We assume that excitons in this system are formed by the electrons located in the one graphene sheet and, correspondingly, the holes located in the other. Since the motion of the electron is restricted in one graphene sheet and the motion of the hole is restricted in the other graphene sheet, we replace the coordinate vectors of the electron and hole by their projections  $\mathbf{r}_1$ and $\mathbf{r}_2$  on plane of one of the graphene sheet. These new in-plane coordinates $\mathbf{r}_1$ and $\mathbf{r}_2$ will be used
everywhere below in our paper. Thus, we reduced the restricted 3D two-body problem to the 2D two-body problem on the graphene plane.
Each honeycomb lattice is characterized by the coordinates $(\mathbf{r}_j,1)$ on sublattice A and $(\mathbf{r}_j,2)$ on sublattice B with $j=1,2$ referring to the two sheets. Then the two-particle wavefunction, describing two particles in different sheets,
reads $\Psi(\mathbf{r}_1,s_1;\mathbf{r}_2,s_2)$, where $\mathbf{r}_1$ and $\mathbf{r}_2$ represent the coordinates of the electron and hole, correspondingly, and $s_1$, $s_2$ are sublattice indices. This wavefunction can also be understood as a four-component spinor, where the spinor components
refer to the four possible values of the sublattice indices $s_1,s_2$;\:
\beq
\label{wavefunction1}
\Psi(\mathbf{r}_1,s_1;\mathbf{r}_2,s_2)=
\left({\begin{array}{c}
\phi_{aa}(\mathbf{r}_1,\mathbf{r}_2) \\
\phi_{ab}(\mathbf{r}_1,\mathbf{r}_2) \\
\phi_{ba}(\mathbf{r}_1,\mathbf{r}_2) \\
\phi_{bb}(\mathbf{r}_1,\mathbf{r}_2)
\end{array}}\right)
\ .
\eeq
In other words, the spinor components are from the same tight-binding wavefunction at different sites.
Each graphene sheet has an energy gap. Obviously the gaps in these sheets are
independent and in the general case we can introduce two non-equal gaps $\delta_1$ and $\delta_2$, corresponding
to the first and the second graphene sheet, respectively. The gap parameters $\delta_1$, $\delta_2$ are the consequence
of adatoms on the graphene sheets, which create a one-particle potential.

The corresponding hopping matrix for two non-interacting particles, including the energy gaps $\delta_1$
and $\delta_2$ on the first and second sheets, correspondingly,  then reads
\begin{equation}
\mathcal{H}_{0}=\left(
\begin{array}{cccc}
-\delta _{1}+\delta _{2} & d_{2} & d_{1} & 0 \\
d_{2}^{\dag} & -\delta _{1}-\delta _{2} & 0 & d_{1} \\
d_{1}^{\dag} & 0 & \delta _{1}+\delta _{2} & d_{2} \\
0 & d_{1}^{\dag} & d_{2}^{\dag} & \delta _{1}-\delta _{2}%
\end{array}%
\right)   \ ,\label{k17}
\end{equation}
In Eq. \eqref{k17} $d_{1} = \hbar v_{F}(-i\frac{\partial }{\partial
x_{1}}-\frac{\partial }{\partial
y_{1}})$, $d_{2}=\hbar v_{F}(-i\frac{\partial }{\partial x_{2}}-\frac{\partial }{%
\partial y_{2}})$ and the corresponding hermitian conjugates are $d_{1}^{\dag}=\hbar v_{F}(-i\frac{\partial }{\partial x_{1}}+\frac{\partial }{%
\partial y_{1}})$, $d_{2}^{\dag}=\hbar v_{F}(-i\frac{\partial }{\partial x_{2}}+\frac{%
\partial }{\partial y_{2}})$, where $x_1$, $y_1$ and $x_2$, $y_2$
are the coordinates of vectors $\mathbf{r}_1$ and $\mathbf{r}_2$, correspondingly,
$v_{F} = \sqrt{3}at/(2\hbar)$  is the Fermi velocity of electrons in
graphene, where $a =2.566 \ \mathrm{\AA}$ is a lattice constant and
$t \approx 2.71 \ \mathrm{eV}$ is the overlap integral between the
nearest carbon atoms \ \cite{Lukose}. This Hamiltonian allows us to write
the eigenvalue equation for two non-interacting  particles as
\bigskip
\begin{equation}
\mathcal{H}_{0}\mathit{\Psi }_{0}=\epsilon_{0}\mathit{\Psi }_{0}
\ , \label{k18}
\end{equation}
which leads to the following eigenenergies:
\begin{equation}
\label{5}
\epsilon_0 (k_1,\delta_1;k_2,\delta_2)=\pm
\sqrt{k_1^2+k_2^2+\delta_1^2+\delta_2^2 \pm 2
\sqrt{(k_1^2+\delta_1^2)(k_2^2+\delta_2^2)}}=\pm\sqrt{k_1^2+\delta_1^2}\pm\sqrt{k_2^2+\delta_2^2} \ .
\end{equation}
where $k_1$ and $k_2$ are  momentum of each particle, correspondingly. Eq.~\eqref{5} gives the energy spectrum for two non-interacting particles in the presents of the non-equal gaps energies $\delta_1$ and $\delta_2$.
The energy dispersion is symmetrical with respect to the replacement of particles $1$ and $2$.  When there are no gaps, $\delta_1=0$ and $\delta_2=0$, as it follows from \eqref{5} the energy dispersion is \;$\pm k_1 \pm k_2$.
\par
Let's now consider the electron and hole located in two graphene sheets with the interlayer separation $D$, and interacting via  the Coulomb potential $V(r) =-e^2/\varepsilon \sqrt{r^2+D^2}$ where $r$ is the projection of
the distance between an electron and a hole on the plane parallel to the graphene sheets, $e$ is the electron charge, and $\epsilon$ is the dielectric constant of the dielectric between graphene sheets. Now the problem for the two
interacting particles located in different graphene sheets with the broken sublattice symmetry in each sheet can be described by the Hamiltonian
\begin{equation}
\mathcal{H}=\left(
\begin{array}{cccc}
-\delta _{1}+\delta _{2}+V(r) & d_{2} & d_{1} & 0 \\
d_{2}^{\dag} & -\delta _{1}-\delta _{2}+V(r) & 0 & d_{1} \\
d_{1}^{\dag} & 0 & \delta _{1}+\delta _{2}+V(r) & d_{2} \\
0 & d_{1}^{\dag} & d_{2}^{\dag} & \delta _{1}-\delta _{2}+V(r)
\end{array}%
\right) \ ,  \label{k20}
\end{equation}
and the eigenvalue problem for Hamiltonian \eqref{k20} is
\begin{equation}
\label{l}
\mathcal{H} \Psi= \epsilon \Psi
\end{equation}
where $\Psi$ are four-component eigenfunctions as given in Eq.\eqref{wavefunction1}.
\par
The Hamiltonian \eqref{k20} describes two interacting particles located in two
graphene sheets and satisfies the following conditions:
\par
i) when the interaction between particles vanished $V(r) = 0$ it describes two
independent particles, each located in the separate graphene sheet, having two
independent gaps energies related to the broken sublattice symmetry in each graphene sheet.
\par
ii) when the gaps in each graphene sheet vanish, $\delta_1 =0$ and $\delta_2 =0$ the Hamiltonian describes two interacting particles in one graphene
 sheet \cite{Sabio} (let us mention that for $\delta_1 = \delta_2 = 0$ and $D = 0$ the Hamiltonian \eqref{k20} is identical to the Hamiltonian $(2)$
 in Ref.~\cite{Sabio} representing the two-particle problem in one graphene graphene sheet if the band gap is absent) if a two-body potential is
$ e^2/\varepsilon r$  or in two graphene sheets with the interlayer separation $D$, and interacting via  the  potential $V (r) =-e^2/\varepsilon
\sqrt{r^2+D^2}$.
\par
iii) when both gaps vanish $\delta_1=0$ and $\delta_2=0$, as well as  two-body potential $V(r)=0$, the Hamiltonian describes two non-interacting Dirac
particles. It is important to mentioned that eigenenergies are symmetrical with respect of replacement particle $1$ and $2$.

\par
In Hamiltonian~(\ref{k20})  the center-of-mass energy can not be
separated from the relative motion even though the interaction
$V=V(r)$ depends only on the coordinate of the relative motion. This is caused by
the chiral nature of Dirac electron in graphene. The similar
conclusion was made for the two-particle problem in graphene in
Ref.~[\onlinecite{Sabio}], where two particles in a single sheet were
considered without gaps and $D=0$.

\section{single exciton eigenvalue problem}

\label{sol}

Since the electron-hole Coulomb interaction depends only on the
relative coordinate, we introduce the new ``center-of-mass''
coordinates in the plane of a graphene sheet $(x,y)$:
\begin{gather}
\label{exp1}
\mathbf{R}=\alpha \mathbf{r}_1+\beta \mathbf{r}_2 \ , \nonumber \\
\mathbf{r} = \mathbf{r}_1-\mathbf{r}_2 \ .
\end{gather}
Here the coefficients $\alpha$ and $\beta$ are to be determined
later. Apparently we can use the analogy of the two-particle problem
for gapped Dirac particles in two-layer graphene with the
center-of-mass coordinates for the case of Schr\"{o}dinger equation.
 The coefficients $\alpha$ and $\beta$
will be found below from the condition of the separation of the
coordinates of the center-of-mass and relative motion in the
Hamiltonian in the one-dimensional ``scalar'' equation determining
the corresponding component of the wave function.

We are  looking for the solution of \eqref{k20} in the form
\begin{equation}
\Psi_j(\mathbf{R},\mathbf{r})=\texttt{e}^{i \mathbf{\mathcal{K}}\cdot
\mathbf{R}} 
\psi_j(\mathbf{r}) \ .
\end{equation}
Let's introduce the following notations:
\begin{gather}
\mathcal{K}_+=\mcKp=\mathcal{K}_x+i \mathcal{K}_y   \ , \nonumber \\
\mathcal{K}_-=\mcKm=\mathcal{K}_x-i \mathcal{K}_y \ , \nonumber \\
\Theta=\tan^{-1}\left({\frac{\mathcal{K}_y}{\mathcal{K}_x}}\right) \ ,
\end{gather}
and rewrite the Hamiltonian 
\eqref{k20} in a form of the $2\times2$ matrix as
\begin{equation}
\label{22ham}
\mathcal{H} = \left({ \begin{array}{cc} \mathcal{O}_2+V(r)\sigma_0-\delta_1\sigma_0+\delta_2\sigma_3 & \mathcal{O}_1 \\
                    \mathcal{O}_1^{\dag} &  \mathcal{O}_2+V(r)\sigma_0-\delta_1\sigma_0+\delta_2\sigma_3
                    \end{array} }\right) \ ,
\end{equation}
where $\mathcal{O}_1$ and $\mathcal{O}_2$ are given by
\begin{gather}
\label{oo}
\mathcal{O}_1= \hbar v_{F} \left({ \alpha \mathcal{K}_{-} - i \p_x - \p_y }\right) \sigma_0=\hbar v_{F} \alpha \mathcal{K}_{-}\sigma_0 - \hbar v_{F} (i \p_x + \p_y)  \sigma_0, \\
\notag
\mathcal{O}_2= \hbar v_{F}\left({ \begin{array}{cc}
                      0  & \beta \mathcal{K}_- + i \p_x + \p_y \\
                      \beta \mathcal{K}_+ + i \p_x - \p_y  & 0
                      \end{array}  }\right)=  \\
                                            \hbar v_{F}\beta \left({ \begin{array}{cc} 0 & \mathcal{K}_x- i \mathcal{K}_y \\
                                                                     \mathcal{K}_x+i \mathcal{K}_y & 0
         \end{array} }\right)+\left({ \begin{array}{cc} 0 & i \p_x + \p_y \\
                                                                     i \p_x - \p_y & 0
         \end{array} }\right) \ ,
                \label{ooo}
\end{gather}
where $x$ and $y$ are the components of vector $\mathbf{r}$, $\sigma_j$  are the  Pauli matrices, $\sigma_0$ is the $2 \times 2$ unit matrix, also $\p_x=\p/\p x$ and $\p_y=\p/\p y$. Analysis of the operators \eqref{oo} and \eqref{ooo} shows that the coordinates of the center-of-mass and relative motion can be separated.

For $\phi_{aa}$ we can rewrite the eigenvalue problem as a one-dimensional equation
(see Appendix \ref{app:A}):
\begin{equation}
\label{fin11} \left({ \frac{(\hbar v_F
\mathcal{K})^2}{2\epsilon}+V(r)-\frac{\epsilon (\hbar v_F)^2
\nabla_\mathbf{r}^2}{2\left({\epsilon^2 -
(\delta_1+\delta_2)^2}\right)}}\right)\phi_{aa}=\left[{\epsilon+\delta_1-\delta_2}\right]
 \phi_{aa} \ .
\end{equation}
The other components of \eqref{wavefunction1} are given as:
\beq
\Psi_b=-(\epsilon\sigma_0-iD_2-\delta_1\sigma_0-\delta_2\sigma_3-V(r)
\sigma_0)^{-1}iD_1^\dagger\Psi_a
\label{eq:8}
\eeq
for
\begin{equation}
\Psi_a=\left({\begin{array}{c}
\phi_{aa} \\
\phi_{ab}
\end{array}}\right) , \ \ \
\Psi_b=\left({\begin{array}{c}
\phi_{ba} \\
\phi_{bb}
\end{array}}\right)
\end{equation}
and
\beq D_1=
\left({ \begin{array}{cc}
\p_{x_1}-i\p_{y_1} & 0 \\
0 & \p_{x_1}-i\p_{y_1} \\
\end{array}  }\right)
=(\p_{x_1}-i\p_{y_1})\sigma_0
\eeq
\beq D_2=\left({ \begin{array}{cc}
0 & \p_{x_2}-i\p_{y_2} \\
\p_{x_2}+i\p_{y_2} & 0 \\
\end{array}}\right)
=\p_{x_2}\sigma_1+\p_{y_2}\sigma_2
\eeq
with Pauli matrices $\sigma_j$ and $2\times2$ unit matrix $\sigma_0$. Moreover, we have
\beq
\phi_{ab}=\left[\epsilon+\delta_1+\delta_2-V(r)+\frac{1}{\epsilon-\delta_1+\delta_2}(\p_{x_1}^2+\p_{y_1}^2)\right]^{-1}(i\p_{x_2}-\p_{y_2})\phi_{aa}
\ .
\label{eq:7}
\eeq
Assuming  $r \ll D$ and substituting the second-order series expansion for the interaction potential
$V(r)=-V_0+\gamma r^2$ into Eq.~(\ref{fin11}), where $V_{0} = e^{2}/(\varepsilon D)$ and $\gamma = e^{2}/(2\varepsilon D^{3})$, we obtain
\begin{equation}
\label{2simpl} \left({-\frac{\epsilon (\hbar v_F)^2
\nabla_\mathbf{r}^2}{2\left({\epsilon^2 - (\delta_1+\delta_2)^2}\right)} +
\gamma
r^2}\right)\phi_{aa}=\left[{\epsilon+\delta_1-\delta_2+V_0-\frac{(\hbar
v_F \mathcal{K})^2}{2\epsilon}}\right] \phi_{aa}
\ .
\end{equation}
The last equation can be rewritten in the form of the two-dimensional isotropic harmonic oscillator:
\begin{equation}
\label{harm}
\left({-\mathcal{F}_1(\epsilon)\nabla_\mathbf{r}^2+\gamma r^2}\right)\phi_{aa}=\mathcal{F}_0(\epsilon) \phi_{aa} \ ,
\end{equation}
where
\begin{gather}
\mathcal{F}_1=\frac{\epsilon (\hbar v_F)^2}{2\left({\epsilon^2 - (\delta_1+\delta_2)^2}\right)}  \ , \nonumber  \\
\mathcal{F}_0=\epsilon+\delta_1-\delta_2+V_0-\frac{(\hbar v_F
\mathcal{K})^2}{2\epsilon} \ .
\end{gather}
The solution of Eq.~(\ref{harm}) is well known (see, for example, Ref.~\cite{arfken85}) and is given by 
\begin{gather}
\label{eqstart}
\frac{\mathcal{F}_0(\epsilon)}{\mathcal{F}_1(\epsilon)}=2N\sqrt{\frac{\gamma}{\mathcal{F}_1(\epsilon)}} \ , \end{gather}
where $N=2n_1+n_2+1$ with
$n_{1} = 0,1,2,3,\ldots $, $n_{2} = 0, \pm 1, \pm 2, \pm 3, \ldots, \pm n_{1}$ are the quantum numbers
of the 2D harmonic oscillator.

\begin{figure}[ht]
\centering
\subfigure[]{
\includegraphics[width=0.3\textwidth]{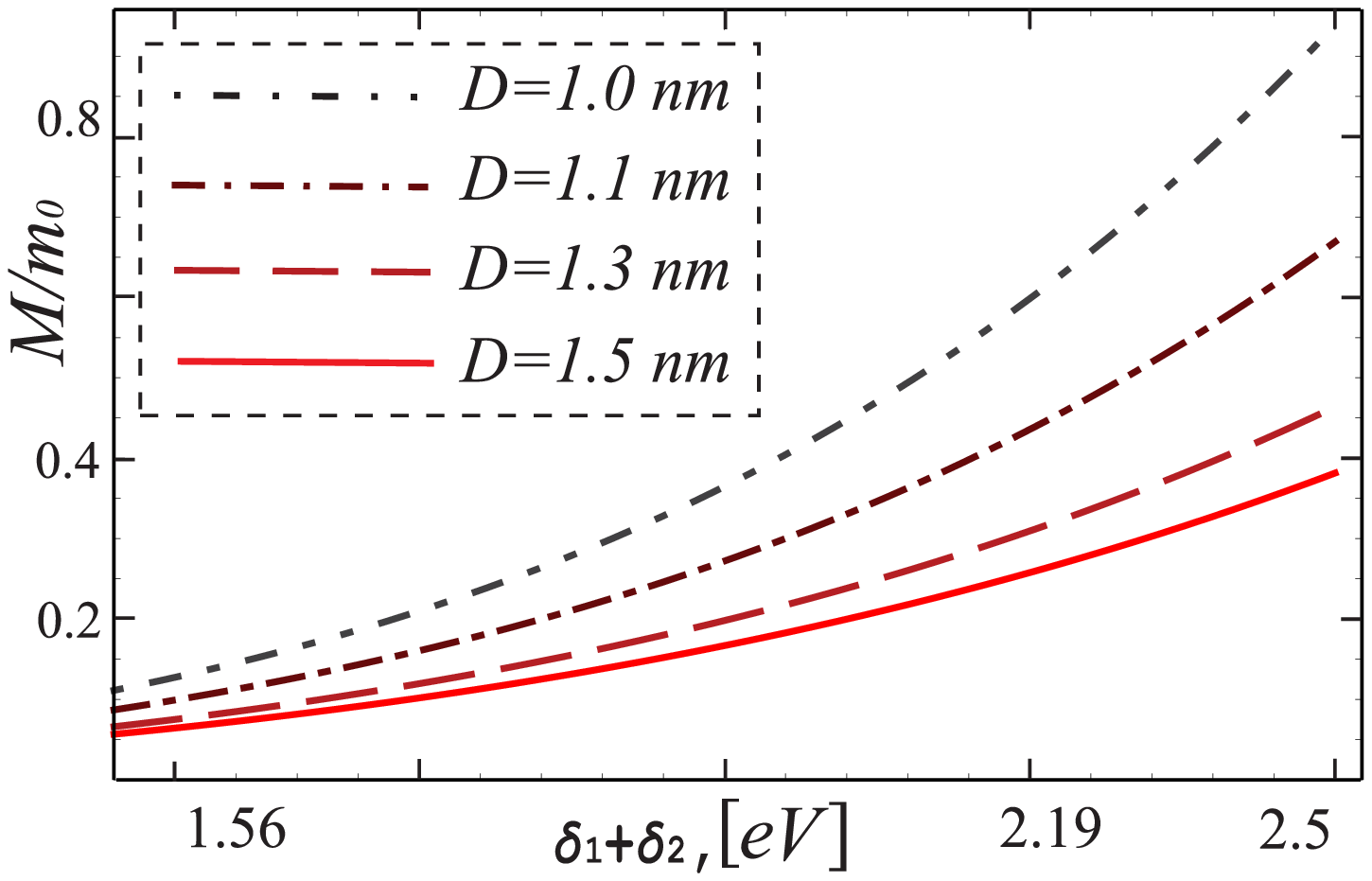}
}
\subfigure[]{
\includegraphics[width=0.3\textwidth]{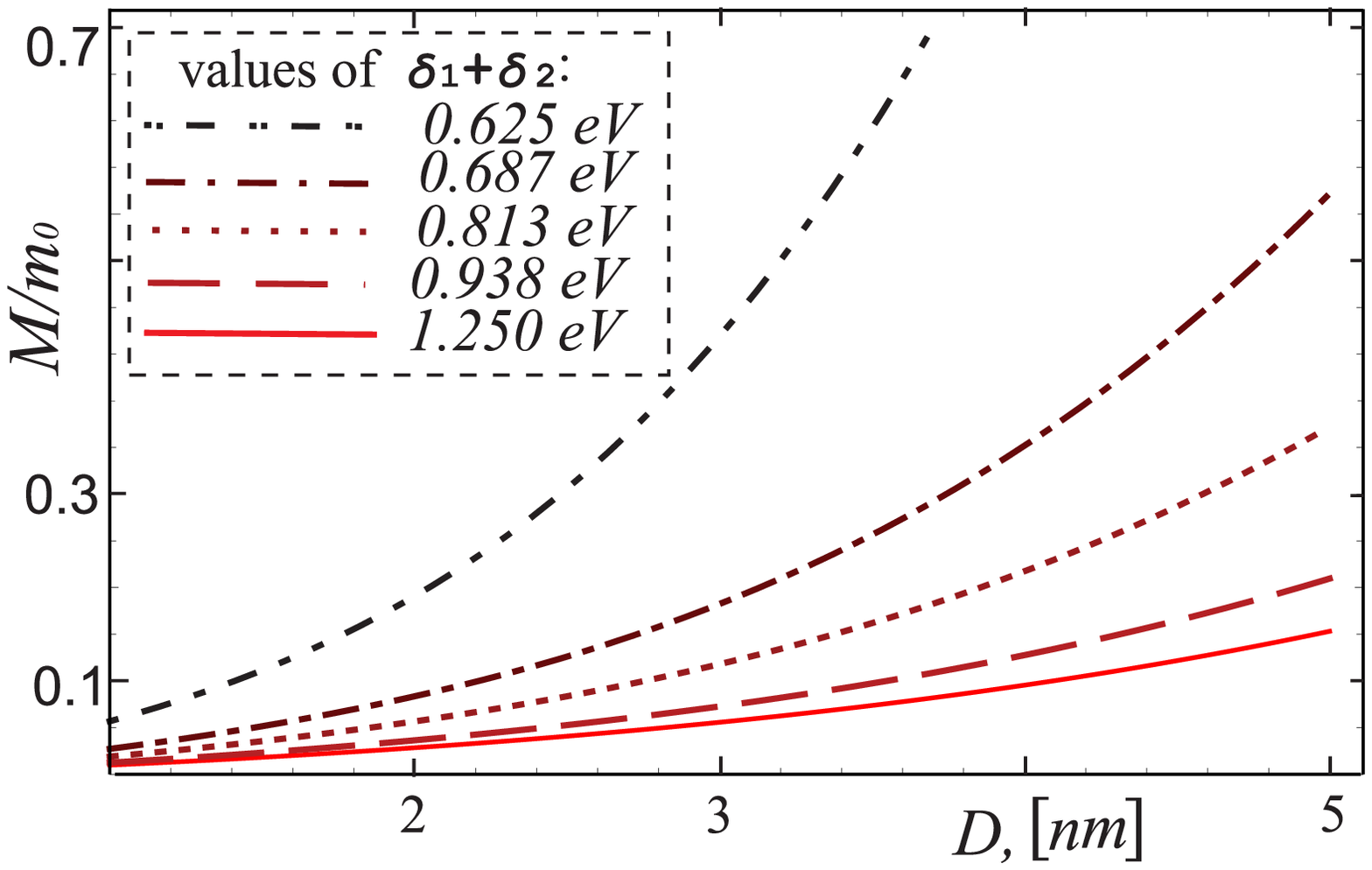}
}
\subfigure[]{
\includegraphics[width=0.28\textwidth]{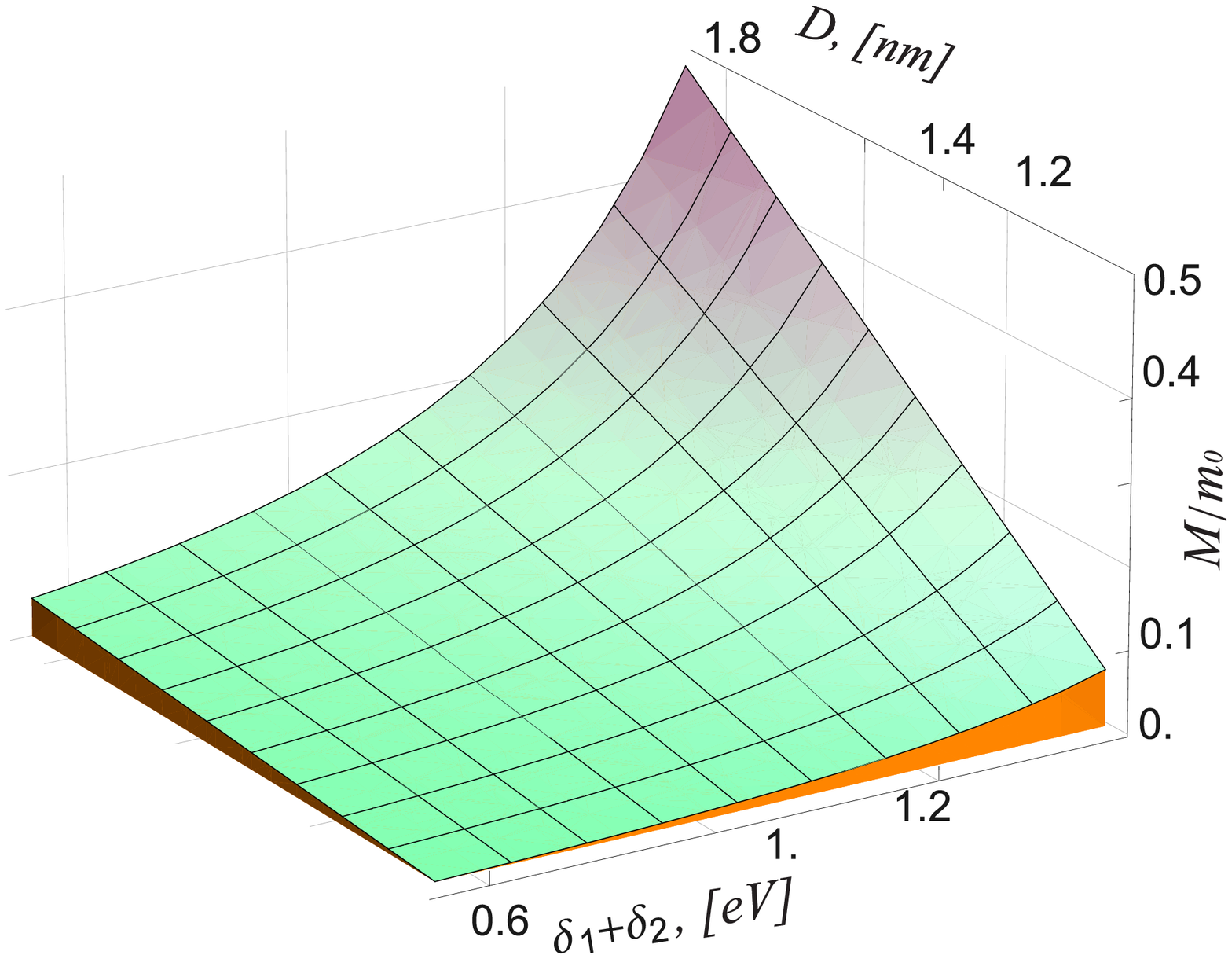}
}
\caption{The effective exciton mass $M$ in the units of free electron mass $m_0$ as a function of the total energy gap for the different graphene interlayer separations $D$ (a), as the function of on interlayer separation $D$ for different values of the total energy gap (b) and as the function of the total energy gap and graphene interlayer separation (c).}
\label{F1}
\end{figure}

After some straightforward but lengthy calculations (cf Appendix \ref{app:B}) we obtain the following expression for the energy in quadratic order with respect to $\mathcal{K}$
\begin{equation}
\label{energy5}
\epsilon=-V_0+\sqrt{\mu^2+\frac{C_1}{\mu}}+\frac{1}{2
\mu^4}\frac{C_1}{\sqrt{1+\frac{C_1}{\mu^3}}}(\hbar v_F
\mathcal{K})^2 \ ,
\end{equation}
where $\mu=\delta_1+\delta_2$ and $C_1=2 \gamma N^2 (\hbar v_F)^2$. Thus, from  \eqref{energy5} we can conclude that the effective exciton mass $M$ is given
as a function of total energy gap $\delta_1+\delta_2$ and the parameter $C_1\propto D^{-3}$ as
\begin{equation}
\label{mass}
M=\frac{\mu^4}{v_{F}^{2}C_1}\sqrt{1+\frac{C_1}{\mu^3}}\ .
\end{equation}
The effective exciton mass $M$ as a function of total energy gap $\delta_1 + \delta_2$ and the interlayer separation D defined by Eq.\eqref{mass} is plotted in Fig.~\ref{F1}. According to Fig.~\ref{F1}, the effective exciton mass $M$ increases when the total energy gap $\delta_1+\delta_2$
and the interlayer separation $D$ increase. The three-dimensional Fig.~\ref{F1}$ c)$ demonstrates dependence of the effective exciton mass  on the total energy gap and interlayer separation.
The dependence of the effective exciton mass $M$  on the interlayer separation $D$ is
caused by the quasi-relativistic Dirac Hamiltonian of
the gapped electrons and holes in graphene layers.  Let us mention that for the excitons
in CQW's the effective exciton mass does not depend on the interlayer separation, because the electrons and holes in CQW's are described by a Schr\"{o}dinger Hamiltonian, while excitons in two graphene layers are described by the Dirac-like Hamiltonian \eqref{k20}.

\section{Collective properties of dipole excitons in a two-layer graphene}
\label{col}

After having found the mass and the energy for a single exciton in the separated double layer of graphene,
we turn now to an ensemble of excitons in this structure.
Due to interlayer separation  $D$ indirect excitons
both in ground state ($n_{1}= n_{2} =0$) and in excited states have non-zero electrical dipole moments.
We assume that indirect exciton interact as {\it parallel} dipoles.
This is valid when $D$ is larger than the mean separation $\langle  r \rangle $
between electron and hole along graphene layers $D \gg \langle  r \rangle$.

\begin{figure}[ht]
\includegraphics[width = 9 cm]{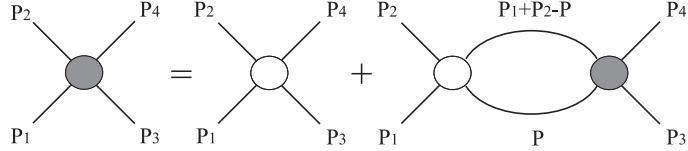}
\caption{The equation for the vertex $\Gamma $ in
momentum representation}
\label{ladder}
\end{figure}

The distinction between exciton and bosons manifests itself in
exchange effects
\cite{Halperin_Rice,KelKoz,Berman,Berman_Willander}. These effects
for exciton with spatially separated electron and hole in a dilute
system $na^{2} \ll 1$ ($n \ll D^{-2}$) are suppressed due to the
negligible overlapping of wave functions of two exciton in the
presence of the potential barrier, associated with the dipole-dipole
repulsion of an indirect exciton \cite{Berman}.  Two indirect
exciton in a dilute system interact as $U(R) =
e^{2}D^{2}/(\varepsilon R^{3})$, where $R$ is the distance between
exciton dipoles along the graphene layers. A small tunneling
parameter $t$ due to this barrier is \cite{Berman_L_G}:
\begin{eqnarray}\label{point}
t=\exp\left[-
\frac{1}{\hbar }\int_{a}^{r_{0}}\sqrt{2M\left(\frac{e^{2}D^{2}}{\varepsilon R^{3}}
- \frac{\kappa ^{2}}{2M}\right)}\
dR\right], \nonumber
\end{eqnarray}
where
\begin{eqnarray}\label{point2}
\kappa ^{2} = \frac{2\pi \hbar^{2}n}{s\log \left(
s\hbar^{4}\varepsilon^{2}/(2\pi n M^2 e^4 D^4) \right)} \nonumber
\end{eqnarray}
is the characteristic value of the center-of-mass exciton momentum
$\hbar \mathcal{K}$ defined as $\kappa=\sqrt{2 M \mu_{ex}}$, where
$\mu_{ex}$ is the chemical potential of the system (see below). In
Eq.~(\ref{point}), $r_{0} = (2M e^{2}D^{2}/\kappa ^{2})^{1/3}$ is
the classical turning point for the dipole-dipole interaction, $s=4$
is the spin degeneracy factor for the excitons and $M$ is the
effective exciton mass in the ground state given by
Eq.~(\ref{mass}). Then the small tunneling parameter $t$ has the
form $t\sim \exp[-2\hbar ^{-1}(M)^{1/2}eD a^{-1/2}]$. Therefore, we
neglect the overlap of the exciton wavefunctions in the limit of
large layer separation $D$ and consider the gas of excitons as a
Bose gas. Consequently, at sufficiently low temperatures the dilute
gas of excitons forms a Bose-Einstein
condensate~\cite{Abrikosov,Griffin}. Formally, the exciton gas can
be treated by the conventional diagram technique for a boson system.
In particular, the effective interaction of the dilute
two-dimensional exciton gas (at $n a^{2} \ll 1$) can be described by
a summation of ladder diagrams \cite{Abrikosov}. From the latter we
obtain an integral equation for vertex function $\Gamma $, depending
on three momenta $\mathbf{p},\mathbf{p}',\mathbf{P}$ and the
frequency $\Omega$, as
\begin{equation}\label{Gamma_Int}
\Gamma (\mathbf{p},\mathbf{p}';\mathbf{P},\Omega) =U (\mathbf{p} - \mathbf{p}')
+ s \int_{}^{} \frac{d^{2} q}{(2\pi \hbar)^2} \frac{U (\mathbf{p} -
\mathbf{q}) \Gamma (\mathbf{q},\mathbf{p}';\mathbf{P},\Omega)}{\frac{\mathcal{L} ^2}{M}
+\Omega - \frac{\mathbf{P}^{2}}{4 M} - \frac{q^2}{M} + i\delta'}
\hspace{0.5cm} (\delta'\rightarrow +0)
\ ,
\end{equation}
where $U(\mathbf{p} - \mathbf{p}')$ is a dipole-dipole interaction
in momentum representation. This equation is also represented by
diagrams in Fig. \ref{ladder}. The chemical potential of the system
is given by
\begin{equation}
\mu_{ex} =\frac{\kappa ^2}{2 M} =  n_{0} \Gamma (0,0;0,0) \equiv n_0 \Gamma_0 \ .
\end{equation}
Equation \eqref{Gamma_Int} can be solved easily when the excitons occupy the ground state
$n_{1} = n_{2} = 0$. Then the energy spectrum of the exciton is given by
$\epsilon(P) = P^{2}/(2M)$, where the mass $M$ is given by Eq. \eqref{mass}.


The specific feature of a two-dimensional
Bose system is connected with the logarithmic divergence of two-dimensional
scattering amplitude at zero energy \cite{Berman,Berman_Willander,Yudson}.
A simple analytical solution of Eq.~(\ref{Gamma_Int}) for the chemical potential can be obtained
if $\kappa M e^{2} D^2/(\hbar^{3}\varepsilon) \ll 1$, which gives for the chemical potential $\mu_{ex}$
\begin{eqnarray}\label{Mu}
\mu_{ex} =  \frac{\kappa ^2 }{2 M}
= \frac{\pi \hbar^{2}n}{s M \log \left[ s\hbar^{4}\varepsilon^{2}/\left(2\pi n M^2 e^4 D^4\right) \right]} \ .
\end{eqnarray}

The solution of Eq.~(\ref{Gamma_Int}) at small momenta provides the
sound spectrum of collective excitations $\epsilon (P)  = c_s P$
with the sound velocity $c_s = \sqrt{\Gamma_0 n/(4 s M)}=
\sqrt{\mu_{ex}/M}$. The appearance of a sound spectrum is a
consequence of the dipole-dipole repulsion. This sound spectrum of
the collective excitations reflects the possibility for the
existence of an excitonic superfluidity at low temperatures in a
double layer graphene, provided that the sound spectrum satisfies to
the Landau criterion of superfluidity \cite{Abrikosov,Griffin}.

\section{Superfluidity of dipole excitons in double layer graphene}
\label{sup}

The dilute exciton gas which was discussed the previous section,
consisting of electron-hole pairs on the graphene double layer,
forms a collective state whose excitations are sound-like modes.
This might be true at low temperatures, whereas at higher
temperatures phase fluctuations can destroy this 2D collective state
by creating vortex-like excitations (i.e. by unbinding
vortex-antivortex pairs). The latter have short-range correlations
which prevent a superfluid state. Therefore, superfluidity is only
possible for temperatures below a critical temperature $T_c$. This
critical temperature describes a Kosterlitz-Thouless
transition~\cite{Kosterlitz} and is defined as
\begin{eqnarray}\label{T_KT}
T_c = \frac{\pi \hbar ^2 n_s (T_c)}{2 k_B M}\ ,
\end{eqnarray}
where $n_s (T)$ is the superfluid density of the exciton system  at the
 temperature $T$, and $k_B$ is Boltzmann constant.

The function $n_s (T)$ in (\ref{T_KT}) can be found from
the relation
$n_s = n - n_n $, where $n$ is the total density and $n_{n}$ is
the normal component density.
We determine the normal component density following the usual procedure
\cite{Abrikosov}. Suppose that the exciton system  moves with
a velocity $\mathbf{u}$. At nonzero temperatures $T$ dissipating quasiparticles
will appear in this system. Since their density is small at low
temperatures, one can assume that the gas of quasiparticles is an ideal
Bose gas. To calculate the superfluid component density
we find the total current of quasiparticles in a
 frame in which the superfluid component is at rest.
Then we obtain the mean total current of 2D excitons in the coordinate system,
moving with a velocity ${\bf u}$:
\begin{eqnarray}
\label{nnor}
\left\langle \mathbf{J} \right\rangle = \frac{1}{M} \left\langle \mathbf{P} \right\rangle =\frac{s}{M}
\int_{}^{}  \frac{d\mathbf{P}}{(2\pi \hbar)^{2}} \mathbf{P}
f\left(\epsilon (P) - \mathbf{P}\mathbf{u} \right)\  ,
\end{eqnarray}
where $f\left(\epsilon (P)\right) = \left(\exp\left[\epsilon (P)/(k_{B}T)\right] - 1\right)^{-1}$ is the Bose-Einstein distribution function.
Expanding the expression inside the integral and leaving the first order by $\mathbf{P}\mathbf{u}/(k_{B}T)$, we have:
\begin{eqnarray}\label{J_Tot}
\langle \mathbf{J} \rangle = - s\frac{\mathbf{u}}{2M}\int\frac{d\mathbf{P}}{(2\pi \hbar)^{2}}P^{2}\frac{\partial f\left(\epsilon (P)\right)}{\partial \epsilon}=
  \frac{3 \zeta (3)s }{2 \pi \hbar^{2}}\frac{k_{B}^{3}T^3}{M c_s^4} \mathbf{u}\ ,
\end{eqnarray}
where $\zeta (z)$ is the Riemann zeta function ($\zeta (3) \simeq
1.202$).
Then we define the
normal component density $n_{n}$ as \cite{Abrikosov}
\begin{eqnarray}\label{J_M}
\langle  \mathbf{J} \rangle = n_n  \mathbf{u}\  .
\end{eqnarray}
Comparing Eqs.~(\ref{J_M}) and~(\ref{J_Tot}), we obtain the expression
for the normal density $n_{n}$, which implies for the superfluid density
\begin{eqnarray}
\label{n_s}
n_s  = n -
 \frac{3 \zeta (3) }{2 \pi \hbar^{2}} \frac{k_{B}^{3}T^3}{c_s^4 M}\ .
\end{eqnarray}
It should be noticed that the expression for the superfluid density
$n_{s}$ of the dilute exciton gas in the double layer graphene in
the presence of the band gaps differs from the corresponding
expression in semiconductor coupled quantum wells (compare with
Refs.~[\onlinecite{Berman,Berman_Willander}] by replacing the total
exciton mass $M = m_{e} + m_h$ with the effective  exciton  mass $M$
given by Eq.~(\ref{mass})).

Using Eq.~(\ref{n_s}) for the density $n_{s}$ of the superfluid component, we obtain an
equation for the Kosterlitz-Thouless transition temperature $T_{c}$. Its solution is
\begin{eqnarray}
\label{tct} T_c = \left[\left( 1 +
\sqrt{\frac{32}{27}\left(\frac{s M k_{B}T_{c}^{0}}{\pi \hbar^{2} n}\right)^{3} +
1} \right)^{1/3}   - \left( \sqrt{\frac{32}{27} \left(\frac{s M k_{B}T_{c}^{0}}{\pi \hbar^{2} n}\right)^{3} + 1} - 1 \right)^{1/3}\right]
\frac{T_{c}^{0}}{ 2^{1/3}} \  ,
\end{eqnarray}
where $T_{c}^{0}$ is the temperature at which the superfluid density vanishes in the
mean-field approximation (i.e., $n_{s}(T_{c}^{0}) = 0$),
\begin{equation}
\label{tct0}
T_c^0 = \frac{1}{k_{B}} \left( \frac{ \pi \hbar^{2} n c_s^4 M }{6 s\zeta (3)}
\right)^{1/3} \ .
\end{equation}

\begin{figure}[h]
\centering
\subfigure[]{
\includegraphics[width=0.32\textwidth]{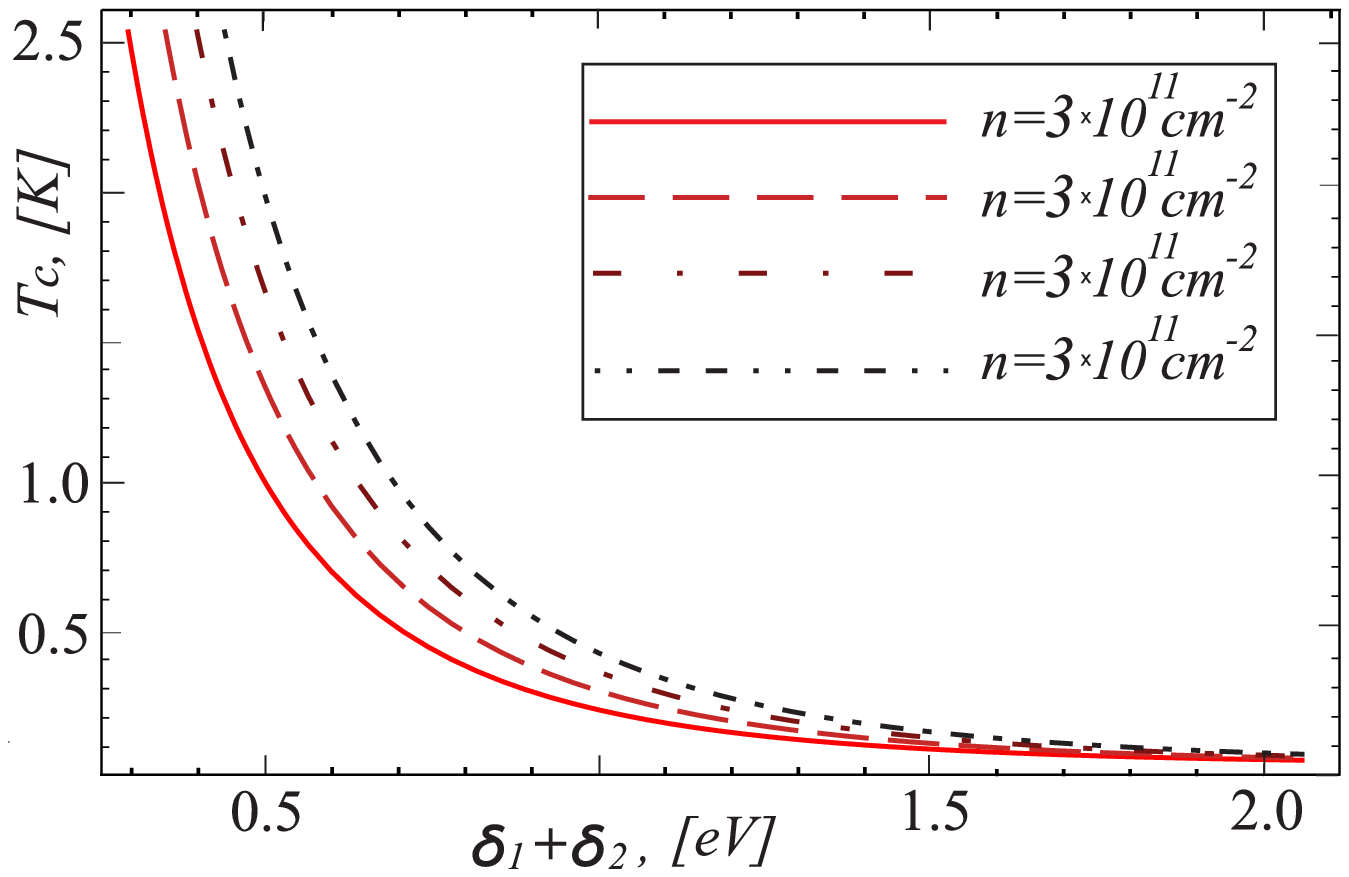}
}
\subfigure[]{
\includegraphics[width=0.31\textwidth]{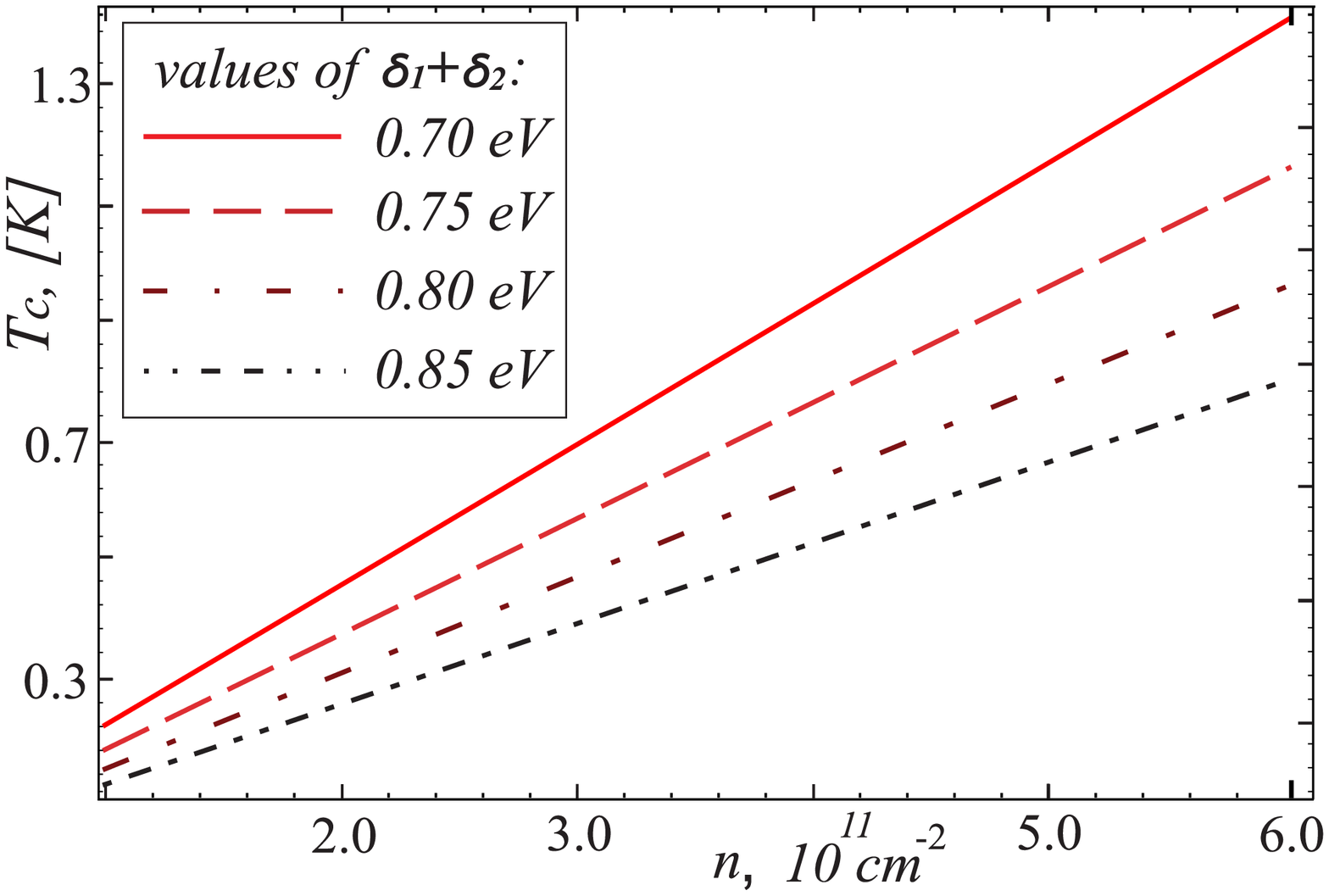}
}
\subfigure[]{
\includegraphics[width=0.29\textwidth]{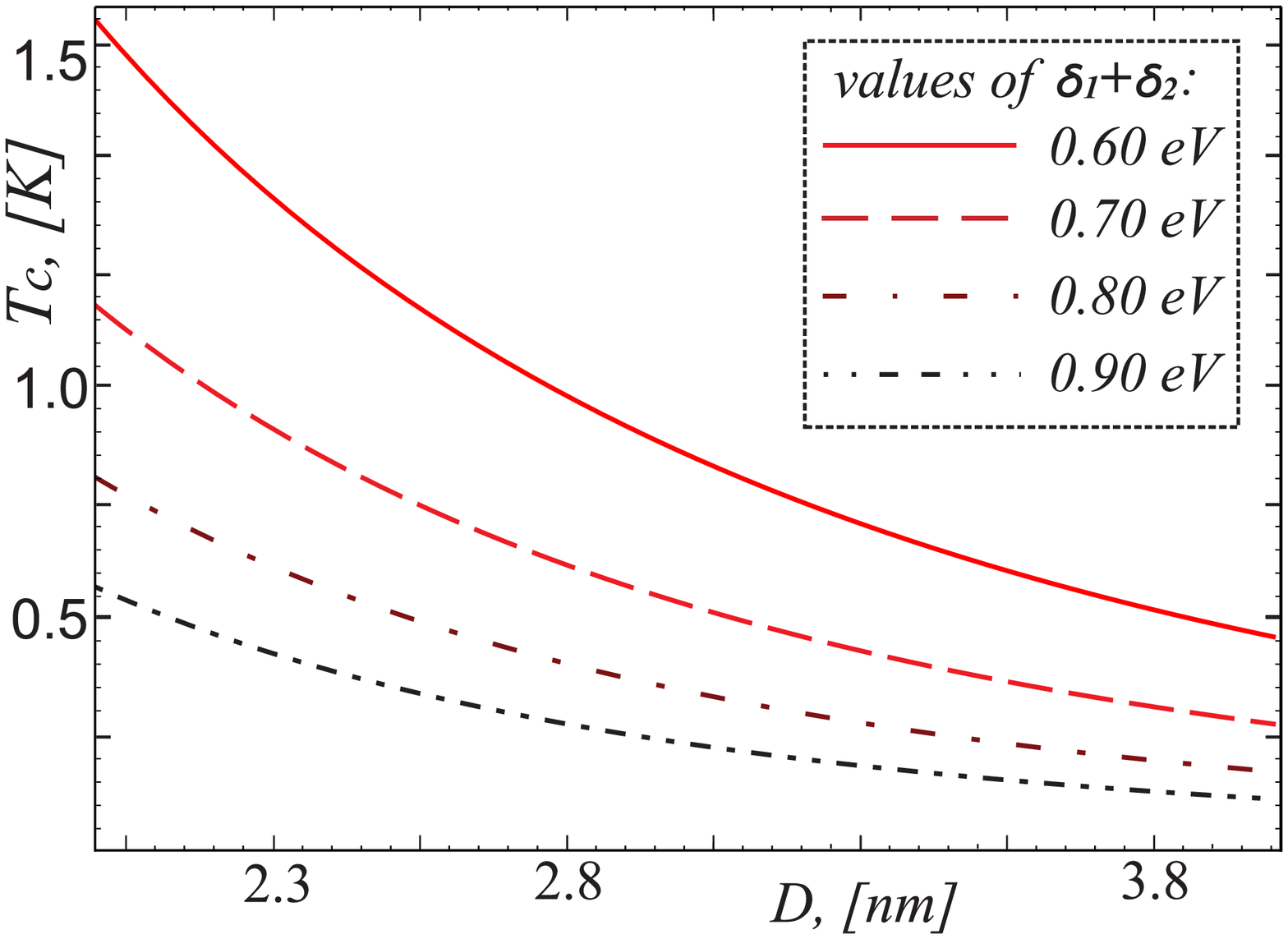}
}

\caption{Kosterlitz-Thouless transition temperature $T_c$ as a function of the total energy gap $\delta_1+\delta_2$, exciton concentration $n$ and the separation between the two graphene layers: (a) demonstrates how the
transition temperature depends on the total energy gap $\delta_1+\delta_2$ for the four different values of exciton mass concentration : $n=3.0\cdot10^{11}cm^{-2};\;4.0\cdot10^{11}cm^{-2};\;5.0\cdot10^{11}cm^{-2};\;6.0\cdot10^{11}cm^{-2}$; (b) shows the transition temperature dependence on the exciton  concentration for ten different values of the total enertgy gap: $(\delta_1+\delta_2=0.70 eV;\;0.75eV;\;0.80 eV;\;0.85 eV)$; (c) exhibits the dependence of the transition temperature on  the interlayer separation $D$ for the different values of the total energy gap values: $(\delta_1+\delta_2=0.60 eV;\;0.70eV;\;0.80 eV;\;0.90 eV)$.
 }
\label{F2}
\end{figure}

The behavior of the Kosterlitz-Thouless transition temperature $T_c$ as a function of the total energy gap, exciton
concentration $n$ and the interlayer separation $D$ is presented in Fig.~\ref{F2},
using Eqs. (\ref{tct}) and (\ref{tct0}).
As we can see in Fig.~\ref{F2}, $T_c$ increases when the exciton concentration $n$
and increases and decreases when total energy gap and and interlayer
separation increaseincreases.

\section{Discussion}
\label{disc}

We have considered an electron-hole pair with attractive Coulomb interaction,
where the electron and the hole live in two different graphene layers separated by dielectric.
The distance between these layers is tunable such that we can vary the strength of the Coulomb
interaction. Moreover, we assume a band gap in the dispersion of the electron
and the hole which is caused, for instance, by doping the graphene layers with
non-carbon atoms. The electron-hole pair forms an exciton whose mass $M$
depends on the sum of the two gaps and on the layer distance $D$, as given by
Eq. \eqref{mass}. This result is generalized to a dilute gas of such excitons,
which experiences a repulsive dipole-dipole interaction. The latter does not pose
a problem because the dipoles are fixed by the double layers and can only interact as parallel
dipoles. This allows us to consider the dilute excitons as point-like bosons which
can be treated in a conventional self-consistent approach for bosons, leading to
an effective interaction which is defined by the integral equation \eqref{Gamma_Int}.
A solution of the latter for point-like particles of mass $M$ provides us a sound-like
spectrum of the quasiparticles, which represents superfluidity.
The advantage of observing the exciton superfluidity and BEC in graphene in comparison with these in CQW's
is based on the fact that the exciton superfluidity and
BEC in graphene can be controlled by the gaps which depend on doping. Note that we considered the superfluidity
in two cases: first, an equilibrium system of electrons and holes created by the gates, and the second case
is the electrons and holes created by the laser pumping such that the excitons are in the quasi-equilibrium thermodynamical state.
 A temperature $T_c$ which is the critical temperature of a Kosterlitz-Thouless
transition was obtained. There is a superfluid state for $T<T_c$ and
a normal state for $T\ge T_c$. The value of this critical
temperature is given by Eq. \eqref{tct}. Using the value of the
total energy gap $\sim 0.26 \ \mathrm{eV}$ from Ref.~\cite{Zhou} and
a interlayer distance $D=10 \ \mathrm{nm}$ we obtain for the
critical temperature $T_c\approx 0.1 \ \mathrm{K}$  for exciton
concentration $n=5\times 10^{11} \ \mathrm{cm^{-2}}$, while for a
interlayer distance $D = 3 \ \mathrm{nm}$ the critical temperature
$T_c \approx 1.3 \ \mathrm{K}$ and for $D = 1 \ \mathrm{nm}$ the
critical temperature becomes $T_c \approx 7.5 \ \mathrm{K}$.

The superfluid
state at $T < T_{c}$ can manifest itself in the existence of
persistent (``superconducting'') electric currents with opposite
directions in the graphene layers. The interlayer tunneling in
 an {\it equilibrium} spatially separated electron-hole system
 leads to interesting
Josephson phenomena in the system: to a transverse Josephson
current, inhomogeneous (many sine-Gordon soliton) longitudinal
  currents, \cite{Klyuchnik}  diamagnetism
for the case of  magnetic field $\mathbf{B}$ parallel to the
junction (when $B$
 is less than a certain critical value $B_{c1}$, depending on the
tunneling coefficient), and a mixed state with Josephson vortices
for $B > B_{c1}$. In addition, taking tunneling into account leads
to the order parameter symmetry breaking and to a change of the
phase transition type. The interlayer resistance relating to the
drag of electrons and holes can also be a sensitive indicator of the
transition to the superfluid state of the electron  hole
system \cite{Vignale_drag,Berman_drag}. The existence of a local superfluid density below $T_{c}$
can be detected, for example, by measuring the characteristic temperature dependence
of the exciton diffusion on intermediate distances \cite{butov96}.

The advantage of observing the exciton superfluidity and BEC in
graphene in comparison with these in CQW's is based on the fact that
the exciton superfluidity and BEC in graphene can be controlled by
the gaps which depend on doping. Note that we considered the
superfluidity in two cases: first, an equilibrium system of
electrons and holes created by the gates, and the second case is the
electrons and holes created by the laser pumping such that the
excitons are in the quasi-equilibrium thermodynamical state.
Another advantage is that graphene is much cleaner than typical
semiconductors used for CQW's, where the roughness of QWs boundaries
is crucial. Therefore, disorder is much less of a problem in double
layer graphene.

In conclusion, we propose a physical realization to observe
Bose-Einstein condensation and superfluidity of
quasi-two-dimensional dipole excitons in two-layer graphene in the
presence of band gaps. The effective exciton mass is calculated
as a function of the electron and the hole energy gaps in the
graphene layers, density and interlayer separation. We demonstrate
the increasing effective exciton mass with the increase of the
gaps and interlayer separation. The dependence of the exciton mass
on the electron-hole Coulomb attraction and interlayer distance
comes from  the Dirac-like spectrum of electrons and holes.  We show that
the superfluid density $n_{s}$ and the Kosterlitz-Thouless
temperature $T_{c}$ increases with increasing excitonic density
$n$ and decreases with the rise of the gaps $\delta_1$ and $\delta_2$, as well as the interlayer
separation $D$, and therefore, could be controlled by these parameters. As we mentioned before, the energy gap
parameters $\delta_1$ and $\delta_2$ are determined by the doping concentration.

\appendix

\section{Eigenvalue Problem for two particles}
\label{app:A}

For the Hamiltonian \eqref{22ham} the eigenvalue problem $\mathcal{H} \Psi = \epsilon \Psi$ results in the following equations:
\begin{eqnarray}
\label{221}
\notag \left({\mathcal{O}_2+V(r)\sigma_0-\delta_1\sigma_0+\delta_2\sigma_3}\right)\Psi_a+\mathcal{O}_1 \Psi_b=\epsilon \sigma_0 \Psi_a \\
\label{222} \mathcal{O}_1^{\dag}\Psi_a +
\left({\mathcal{O}_2+V(r)\sigma_0-\delta_1\sigma_0+\delta_2\sigma_3}\right)
\Psi_b=\epsilon \sigma_0 \Psi_b \ .
\end{eqnarray}
From Eq.~\eqref{221} we have:
\begin{equation}
\label{psi1}
\Psi_b=\left({\epsilon\sigma_0-\mathcal{O}_2-V(r)\sigma_0+\delta_1\sigma_0-\delta_2\sigma_3}\right)^{-1}\mathcal{O}_1^{\dag}\Psi_a
\ .
\end{equation}
Assuming the interaction potential and both relative and
center-of-mass kinetic energies are small compared to the gaps
$\delta_1$ and $\delta_2$ we use the following approximation:
\begin{equation}
\label{apr}
\left({\epsilon \sigma_0-\mathcal{O}_2-V(r)\sigma_0+\delta_1\sigma_0-\delta_2\sigma_3}\right)^{-1}
\backsimeq \frac{1}{\epsilon\sigma_0+\delta_1\sigma_0-\delta_2\sigma_3} \ .
\end{equation}
 Using the fact that the operator $\mathcal{O}_1^{\dag} \mathcal{O}_1$ is purely
hermitian, applying Eq.~\eqref{221} and
\begin{equation}
\mathcal{O}_1^{\dag}\mathcal{O}_1= \hbar^{2}v_{F}^{2}
\left({\alpha^2 \mathcal{K}^2-\nabla_\mathbf{r}^2-2 i \alpha (\mathcal{K}_{x}
\p_y+\mathcal{K}_{y} \p_x)}\right)\sigma_0 \ ,
\end{equation}
we obtain:
\begin{equation}
\label{19}
\left({\mathcal{O}_2+V(r)\sigma_0-\delta_1\sigma_0+\delta_2\sigma_3}\right)\Psi_a+
\hbar^{2}v_{F}^{2} \frac{\left({\alpha^2 \mathcal{K}^2-\nabla_\mathbf{r}^2-2
i \alpha (\mathcal{K}_{x} \p_x+\mathcal{K}_{y} \p_y)}\right)}{\epsilon\sigma_0+\delta_1\sigma_0-\delta_2\sigma_3}\Psi_a=\epsilon\sigma_0
\Psi_a \ .
\end{equation}

Now we rewrite Eq.~\eqref{19} in the following form:
\begin{gather}
\label{231}
\left({ -\delta_1+\delta_2+V(r)+ \hbar^{2}v_{F}^{2}\frac{\alpha^2 \mathcal{K}^2-\nabla_\mathbf{r}^2-2 i \hbar v_{F}\alpha (\mathcal{K}_{x}
\p_x+\mathcal{K}_{y} \p_y)}{\epsilon - \delta_1-\delta_2} }\right)\phi_{aa}+ \\
\notag
\hbar v_{F}\left({\beta \mathcal{K}_- + i \p_x + \p_y}\right)
\phi_{ab}=\epsilon \phi_{aa} \ , \\
\label{232}\hbar v_{F} \left({ \beta \mathcal{K}_+ + i \p_x - \p_y}\right)\phi_{aa}+ \\
\left({-\delta_1-\delta_2+V(r)+
\hbar^{2}v_{F}^{2}\frac{\alpha^2 \mathcal{K}^2-\nabla_\mathbf{r}^2-2 i \alpha
(\mathcal{K}_{x} \p_x +\mathcal{K}_{y} \p_y)}{\epsilon -
\delta_1+\delta_2}}\right) \phi_{ab}=
\epsilon \phi_{ab} \ .
\end{gather}
We solve Eq.~\eqref{232} with respect to $\psi_{ab}$:
\begin{equation}
\label{ab}
\psi_{ab}=\left[{\epsilon+\delta_1+\delta_2-V(r)-\hbar^{2}v_{F}^{2}\frac{\alpha^2
\mathcal{K}^2-\nabla_\mathbf{r}^2-2 i \alpha (\mathcal{K}_{x}
\p_x+\mathcal{K}_{y} \p_y)}{\epsilon -
\delta_1+\delta_2}}\right]^{-1}\left({ \beta \mathcal{K}_+ + i \p_x - \p_y}\right)\hbar v_{F}\psi_{aa} \ .
\end{equation}
Substituting  $\psi_{ab}$ from Eq.~(\ref{ab}) into Eq.~(\ref{231}),
we obtain:
\begin{eqnarray}
\label{long} \notag & & \left({
-\delta_1+\delta_2+V(r)+\hbar^{2}v_{F}^{2}\frac{\alpha^2
\mathcal{K}^2-\nabla_\mathbf{r}^2-2 i \alpha (\mathcal{K}_{x} \p_x+\mathcal{K}_{y} \p_y)}{\epsilon - \delta_1-\delta_2} }\right)\phi_{aa}+ \\
\notag
&+& \hbar^{2}v_{F}^{2}\left({\beta \mathcal{K}_- + i \p_x + \p_y}\right)
\left[{\epsilon+\delta_1+\delta_2-V(r)-\hbar^{2}v_{F}^{2}\frac{\alpha^2
\mathcal{K}^2-\nabla_\mathbf{r}^2-2 i \alpha (\mathcal{K}_{x} \p_x+\mathcal{K}_{y} \p
_y)}{\epsilon - \delta_1+\delta_2}}\right]^{-1} \\
&\times&\left({ \beta \mathcal{K}_+ + i \p_x - \p_y}\right)=\epsilon
\phi_{aa} \ .
\end{eqnarray}
Assuming again that the interaction potential and both relative and
center-of-mass kinetic energies are small compared to the gaps
$\delta_1$ and $\delta_2$ we apply to Eq.~(\ref{long}) the following
approximation:
\begin{equation}
\label{appr} \left[{\epsilon+\delta_1+\delta_2-V(r)-
\hbar^{2}v_{F}^{2}\frac{\alpha^2 \mathcal{K}^2-\nabla_\mathbf{r}^2-2 i \alpha
(\mathcal{K}_{x} \p_x+\mathcal{K}_{y} \p_y)}{\epsilon -
\delta_1+\delta_2}}\right]^{-1}=\frac{1}{\epsilon+\delta_1+\delta_2}
\  .
\end{equation}
Applying the approximation given by Eq.~(\ref{appr}) to
Eq.~(\ref{long}), we get from Eq.~(\ref{long}) the eigenvalue
equation in the form:
\begin{widetext}
\begin{gather}
\notag
\left({ -\delta_1+\delta_2+V(r)+
(\hbar v_{F})^{2}\frac{\alpha^2 \mathcal{K}^2-\nabla_\mathbf{r}^2-2 i \alpha
(\mathcal{K}_{x} \p_x+\mathcal{K}_{y} \p_y)}{\epsilon -
\delta_1-\delta_2}+(\hbar v_{F})^{2}\frac{\beta^2
\mathcal{K}^2-\nabla_\mathbf{r}^2 + 2 i \beta (\mathcal{K}_{x} \p_x+\mathcal{K}_{y} \p_y)}{\epsilon +
\delta_1+\delta_2}}\right)\phi_{aa}    \\
=\epsilon \phi_{aa} \ .
\label{fin1}
\end{gather}
\end{widetext}


Choosing the values for the coefficients  $\alpha$ and $\beta$ to
separate the coordinates of the center-of-mass (the wave vector
$\mathbf{\mathcal{K}}$) and the coordinates relative motion
$\mathbf{r}$ in the Hamiltonian in the l.h.s. of Eq.~(\ref{fin1}),
we have
\begin{eqnarray}
\label{albet}
\alpha=\frac{\epsilon-\delta_1-\delta_2}{2 \epsilon}  \ , \nonumber \\
\beta=\frac{\epsilon+\delta_1+\delta_2}{2\epsilon} \ .
\end{eqnarray}
Substitution of Eq.~(\ref{albet}) into Eq.~(\ref{fin1}) results in Eq. \eqref{fin11}.

\section{Energy spectrum of an exciton}
\label{app:B}

Squaring both sides of Eq. \eqref{eqstart}, we get
\begin{gather}
\label{step2}
\mathcal{F}_0^2=4 N^2 \gamma \mathcal{F}_1 \ .
\end{gather}

Let us introduce the following notations:
\begin{gather}
\mu=\delta_1+\delta_2  \ , \nonumber \\
\nu=\delta_1-\delta_2+ V_0  \ , \nonumber \\
C_1=2 \gamma N^2 (\hbar v_F)^2  \ ,
\end{gather}
which allows us to rewrite the Eq.\eqref{step2} in the form:
\begin{equation}
\label{ineq} \left(\epsilon+\nu-\frac{(\hbar v_F
\mathcal{K})^2}{2\epsilon}\right)^2=C_1
\frac{\epsilon}{\epsilon^2-\mu^2} \ .
\end{equation}
We can rewrite Eq.~(\ref{ineq}) as the form of the equation for $\epsilon$:
\begin{gather}
\label{final1}
\epsilon^5+A \epsilon^4+ B \epsilon^3 + C \epsilon^2+D \epsilon + G = 0
\end{gather}
with the coefficients:
\begin{gather}
A=-2\nu  \ , \nonumber \\
B= \nu^2-(\hbar v_F \mathcal{K})^2-\mu^2 \ , \nonumber \\
C= (2 \mu^2 - (\hbar v_F \mathcal{K})^2)\nu -C_1 \ , \nonumber \\
D= ((\hbar v_F \mathcal{K})^2-\nu^2)\mu^2 \ , \nonumber \\
G=  (\hbar v_F \mathcal{K})^2 \mu^2 \nu \ .
\end{gather}
If $\nu=0\;(\delta_1=\delta_2\;\;\&\;\;\epsilon=-V_0+\epsilon')$ then Eq. \eqref{final1} has the form:
\begin{equation}
\label{analyticalEq} (\epsilon^2-(\hbar v_F
\mathcal{K})^2)(\epsilon^2-\mu^2)-C_1 \epsilon=0
\end{equation}
with $C_1 \ll \epsilon (\epsilon^{2} - \mu^{2})$ and $\hbar v_F
\mathcal{K} \ll \mu$.
\par
First, we assume $C_1=0$ and obtain $\epsilon_0 = \pm \hbar v_F K$ and $\epsilon_0 = \pm \mu$.
\begin{gather}
\label{lat1}
\epsilon_0=\pm \hbar V_F \mathcal{K} \\
\zeta = \epsilon^2 = (\hbar V_F \mathcal{K})^2 + \Delta  \ ,
\end{gather}
where $\Delta$ is very small correction.

We substitute Eq.~(\ref{lat1}) to Eq.~\eqref{analyticalEq} and neglect all the higher order quantities with respect to $\Delta$:
\begin{equation}
 \epsilon=-V_0+\sqrt{(\hbar v_F \mathcal{K})^2+\Delta^2} \backsim -V_0 + \sqrt{\frac{-C_1 \hbar v_F \mathcal{K}}{\mu^2}} \ ,
 \end{equation}
which is not real, and, therefore, does not correspond to physical reality.

Now let's consider the second solution of the zero-order expansion of Eq.~(\ref{analyticalEq}): $\epsilon_0=\pm \mu$. Following the similar procedure, we obtain:
\begin{gather}
\label{2eq} (\epsilon^2-(\hbar v_F
\mathcal{K})^2)(\epsilon^2-\mu^2)-C_1\epsilon = 0  \ .
\end{gather}
Now we substitute $\zeta=\epsilon^2=\mu^2+\Delta$ that into Eq.\eqref{2eq}, where $\Delta$ is the small correction:
\begin{equation}
\Delta\left({\mu^2-(\hbar v_F
\mathcal{K})^2}\right)-C_1\sqrt{\mu^2+\Delta}=0 \ ,
\end{equation}
and solve for $\Delta$ neglecting higher orders with respect to $\Delta$:
\begin{equation}
\Delta=\frac{C_1 \mu}{\mu^2-(\hbar v_F
\mathcal{K})^2}=\frac{C_1}{\mu}(1-\frac{(\hbar v_F
\mathcal{K})^2}{\mu^2})^{-1} \backsim \frac{C_1}{\mu^3}(\mu^2+(\hbar
v_F \mathcal{K})^2) \ ,
\end{equation}
which results in the following energy dispersion:
\begin{equation}
\label{series}
\epsilon=-V_0+\sqrt{\mu^2+\Delta}=-V_0+\sqrt{\mu^2+\frac{C_1}{\mu^3}(\mu^2+(\hbar
v_F
\mathcal{K})^2)}=-V_0+\sqrt{\left({\mu^2+\frac{C_1}{\mu}}\right)+\frac{C_1}{\mu^3}(\hbar
v_F \mathcal{K})^2)} \ .
\end{equation}
We expand this in powers of $\mathcal{K}$ and obtain in second order
\begin{equation}
\epsilon=-V_0+\sqrt{\mu^2+\frac{C_1}{\mu}}\cdot \left({1+\frac{C_1
(\hbar v_F \mathcal{K})^2}{\mu^3
\left({\mu^2+\frac{C_1}{\mu}}\right)}}\right) \ .
\end{equation}
This leads to Eq. \eqref{energy5}.

\end{document}